# DC-conductivity of a suspension of insulating particles with internal rotation

N. Pannacci, E. Lemaire and L. Lobry

Laboratoire de Physique de la Matière Condensée
CNRS Université de Nice-Sophia Antipolis
Parc Valrose, 06108 Nice cedex 2, France

*Abstract* We analyse the consequences of Quincke rotation on the conductivity of a suspension. Quincke rotation refers to the spontaneous rotation of insulating particles dispersed in a slightly conducting liquid and subject to a high DC electric field: above a critical field, each particle rotates continuously around itself with an axis pointing in any direction perpendicular to the DC field. When the suspension is subject to an electric field lower than the threshold one, the presence of insulating particles in the host liquid decreases the bulk conductivity since the particles form obstacles to the ions migration. But for electric fields higher than the critical one, the particles rotate and facilitate the ion migration: the effective conductivity of the suspension is increased. We provide a theoretical analysis of the impact of Quincke rotation on the apparent conductivity of a suspension and we present experimental results obtained with a suspension of PMMA particles dispersed in weakly conducting liquids.

## 1. Introduction

When a non conducting particle immersed into a low conducting liquid is submitted to a sufficiently high DC field, it can rotate spontaneously around itself along any axis perpendicular to the electric field. This symmetry break is known as Quincke rotation from the name of the man who first observed it at the end of the 19[th] century [1]. As explained by Secker and Scialom [2] and later by Melcher [3], the mechanism responsible for Quincke rotation deals with the action of free electrical charges which are contained in the liquid. Indeed, when the electric field, $\mathbf{E_0}$, is applied, the charges contained both in the liquid and in the particle migrate under coulombic force. Then, as stated by Jones [4], depending on the relative magnitude of the charge relaxation times in the liquid and in the particle, the charge distribution that builds at the particle/liquid interface is equivalent to a dipole, $\mathbf{P}$, which is either in the direction of the field or in the opposite direction. In the following, we will call $\mathbf{P}$ the retarded dipole since, as it is associated with the charge distribution at the particle/liquid interface, it evolves with a finite characteristic time which is the Maxwell time, $\tau$. The charge relaxation times in the liquid, $\tau_1$, and in the particle, $\tau_2$, are given by the ratio of the permittivity, $\varepsilon_i$, to the conductivity, $\gamma_i$, i=1,2 refers to the liquid and the particle respectively. When $\tau_2 < \tau_1$, the induced dipole is along the electric field direction and the configuration is stable. In the opposite case, i.e. $\tau_2 > \tau_1$, the dipole, $\mathbf{P}$, which has been created by the accumulation of the charges at the particle/liquid interface is opposite to the electric field direction (Figure 1.a). In this last case, if the particle is slightly rotated, the deviation of its dipole moment $\mathbf{P}$ produces a torque, $\mathbf{\Gamma^E} = \mathbf{P} \times \mathbf{E_0}$, which tends to increase the angular tilt further (Figure 1.b). So, if the electric field intensity is high enough for the electric torque to balance the viscous resistant torque exerted by the surrounding liquid on the particle, the particle will rotate continuously around itself with an axis pointing in any direction perpendicular to the DC electric field.

Quincke rotation can have many implications on the behaviour of a suspension. In a previous paper, we have shown that it was possible to reduce significantly (by about one order of magnitude) the effective viscosity of such a suspension. Indeed when the suspension is motionless and a DC field is applied, particles start rotating around themselves in any direction perpendicular to the field and so the average spin rate of the suspension is zero. But, when a velocity gradient is applied along the field direction, the particles rotation axes will be favoured in the vorticity direction. Therefore, the degeneracy of the rotation direction will be lifted giving rise to a nonzero spin rate of the ensemble of particles. This macroscopic spin rate drives the suspending liquid and thus leads to a decrease of the apparent viscosity of the suspension [5]. Quincke rotation has also interesting consequences on the apparent conductivity of a suspension. In a previous paper [6], we have shown that the particle electro-rotation was responsible for a sensitive increase of the electrical conductivity. Depending on the physical parameters of the liquid and of the particles, the conductivity enhancement induced by the electro-rotation can be large enough for the average suspension conductivity to be higher than that of the suspending liquid whereas the particles are insulating. After having demonstrated this effect with an example [6], we would like to give further experimental details which may be of interest in electrohydrodynamics and to discuss the role played by the suspension organisation and the particle interactions on their rotation. In the following section, we briefly recall how to obtain the dipole moment, **P**, of each rotating particle through a relaxation equation and we present the derivation of the electric current passing through the suspension of rotating particles submitted to the DC electric field. The third part is dedicated to a detailed presentation of the experimental procedure that we use to measure very accurately the conductivity of the suspension. The results obtained with two suspensions made of PMMA particles dispersed in two different dielectric oils are presented and discussed in the fourth section wherein the role played by the suspension organisation is highlighted.

## 2. Theory

### *2.1. Quincke rotation*

Let us consider a spherical particle with dielectric loss (conductivity $\gamma_2$ and permittivity $\varepsilon_2$) surrounded by a lossy liquid (conductivity $\gamma_1$ and permittivity $\varepsilon_1$). When submitted to a DC electric field $\mathbf{E_0}=E_0\mathbf{e_z}$ (Figure 1), the particle acquires a dipole moment that is the sum of two components: the instantaneous dipole moment, $\mathbf{P}^\infty$, coming from the permittivity mismatch between the particle and the liquid:

$$\mathbf{P}^\infty = 4\pi\varepsilon_1 a^3 \beta^\infty \mathbf{E_0} = 4\pi\varepsilon_1 a^3 \frac{\varepsilon_2 - \varepsilon_1}{\varepsilon_2 + 2\varepsilon_1} \mathbf{E_0} \tag{1}$$

and a retarded dipole moment, **P**, whose characteristic time to appear is the Maxwell Wagner time:

$$\tau = \frac{\varepsilon_2 + 2\varepsilon_1}{\gamma_2 + 2\gamma_1} \tag{2}$$

By means of the usual boundary conditions for the electric field, the relation between the retarded dipole moment and the free electric charge distribution, $\sigma_v$, at the particle/liquid interface is obtained:

$$\sigma_v = \frac{2\varepsilon_1 + \varepsilon_2}{4\pi\varepsilon_1 a^3} \mathbf{P}.\mathbf{e_r} \tag{3}$$

Then, performing a charge balance at the surface of the particle which is rotating, a relaxation equation for the retarded part of the dipole moment is straightforwardly obtained [6], [7]:

$$\frac{\partial \mathbf{P}}{\partial t} = (\boldsymbol{\omega} \times \mathbf{P}) - \frac{1}{\tau}\left(\mathbf{P} - 4\pi\varepsilon_1 a^3(\beta^0 - \beta^\infty)\mathbf{E_0}\right) \tag{4}$$

where $\beta^0$ is the Clausius-Mossotti factor of a spherical particle at low frequency: $\beta^0 = \dfrac{\gamma_2 - \gamma_1}{\gamma_2 + 2\gamma_1}$.

The stationary components of the retarded dipole moment are therefore:

$$\mathbf{P}(X_0) = 4\pi\varepsilon_1 a^3 E_0 \begin{pmatrix} 0 \\ -\dfrac{X_0(\beta^0 - \beta^\infty)}{1+X^2} \\ \dfrac{(\beta^0 - \beta^\infty)}{1+X_0^2} \end{pmatrix} \qquad (5)$$

where $X_0$ is the reduced angular velocity of the particle, $X_0 = \omega\tau$ whose stationary value is obtained upon solving the torque balance equation:

$$\mathbf{\Gamma^E} + \mathbf{\Gamma^H} = 0 \qquad (6)$$

$\mathbf{\Gamma^H} = -\alpha\omega$ is the resistant viscous torque. In this expression the rotational friction coefficient $\alpha$ is given by $\alpha = 8\pi\eta a^3$ where $\eta$ is the viscosity of the liquid. $\mathbf{\Gamma^E}$ is the electric torque exerted by the electric field on the charge distribution stuck at the particle surface:

$$\mathbf{\Gamma^E} = \mathbf{P} \times \mathbf{E_0} \qquad (7)$$

The stationary angular velocity of a particle under the action of a DC electric field is then obtained:

$$\begin{array}{l} X_0 = 0 \ ; \ E_0 < E_c \\ X_0 = \pm\sqrt{\left(\dfrac{E_0}{E_c}\right)^2 - 1} \ ; \ E_0 > E_c \end{array} \quad \text{with } E_c = \sqrt{\dfrac{-2\eta}{\varepsilon_1\tau\left(\beta^0 - \beta^\infty\right)}} \qquad (8)$$

It should be stressed that the angular velocity does not depend on the particle size and above the critical field intensity, $E_c$, the particle is expected to rotate around itself with an axis pointing in any direction perpendicular to the field. This last remark is important for the following since the behaviour of an ensemble of particles will be considered. We shall make the assumption that, in the plane perpendicular to the field direction, no rotation axis is favoured so that $<X_0>=0$ and $<P_y>=0$ (the brackets denoting the average over all the particles belonging to the suspension).

### *2.2. Effective conductivity of an ensemble of rotating particles.*

In a previous paper [6], we have shown that the electric current crossing a suspension whose particles are rotating is given by the sum of the bulk current passing through the solid and the liquid phases, $\mathbf{j_c}$ and a surface current, $\mathbf{j_S}$, coming from the convection of the interfacial charges by the particles' rotation:

$$\mathbf{j_c} = (1-\phi)\gamma_1 \langle \mathbf{E_1} \rangle + \phi\gamma_2 \langle \mathbf{E_2} \rangle \tag{9}$$

$$\mathbf{j_s} = \frac{2\varepsilon_1 + \varepsilon_2}{4\pi\varepsilon_1 a^3} \phi \boldsymbol{\omega} \times \mathbf{P} \tag{10}$$

where $\phi$ is the particle volume fraction. $<\mathbf{E_1}>$ denotes the average electric field in the suspending liquid while the brackets around $\mathbf{E_2}$ denote that the field is averaged over all the rotating particles in the suspension. In particular this averaging accounts for the isotropic distribution of the particle rotation axes in the plane perpendicular to the external field direction and results in vanishing the transverse electric field ($<E_{2y}>=0$). $\mathbf{E_1}$ and $\mathbf{E_2}$ are related to each other and to the applied field $\mathbf{E_0}$ by:

$$\mathbf{E_0} = (1-\phi)\langle \mathbf{E_1} \rangle + \phi \langle \mathbf{E_2} \rangle \tag{11}$$

$$\mathbf{E_2} = \mathbf{E_0} - \frac{\mathbf{P}^\infty + \mathbf{P}}{4\pi\varepsilon_1 a^3} \tag{12}$$

Summing (9) and (10) and expressing $\mathbf{E_1}$, $\mathbf{E_2}$, $\mathbf{P}^\infty$ and $\mathbf{P}$ as a function of $\mathbf{E_0}$, the total current density is obtained and the effective conductivity, $\gamma_{s0}$, is deduced by dividing $J_z$ by the external electric field intensity:

$$\gamma_{s0} = \gamma_1 (1 + 3A_0 \phi) \quad \text{with} \quad A_0 = \frac{\beta^0 + \beta^\infty X_0^2}{1 + X_0^2} \tag{13}$$

where $X_0$ is given by eq. (8).

This derivation of the effective conductivity which has been already presented and discussed in [6] does not account for particle interaction effects. It has been already shown [9], [10] that the mutual interaction of two neighbouring rotating particles is responsible for a decrease of the transverse component of their dipole moment and, in turn, of their spin rate. In particular the authors show that the critical field beyond which particles start to rotate is increased by the presence of a neighbouring particle [9] and that the particle angular velocity is reduced to:

$$X = \sqrt{\left(\frac{E_0}{E_c}\right)^2 - (1-G(r))^2} \qquad (14)$$

where G(r) accounts for the multipolar interactions between a pair of particles and is a negative increasing function of their separation distance.

Another way to take into account the interactions between particles is to compute the effective electric field to which a particle belonging to the suspension is subjected. Namely, this effective field is the average electric field in the suspending liquid, $\langle \mathbf{E_1} \rangle$. The field inside a particle is then written as:

$$\mathbf{E_2} = \langle \mathbf{E_1} \rangle - \frac{\mathbf{P}^\infty(\langle \mathbf{E_1} \rangle) + \mathbf{P}(\langle \mathbf{E_1} \rangle)}{4\pi\varepsilon_1 a^3} \qquad (15)$$

Then, using equations (5), (11) and (15), the effective electric field is obtained in a self-consistent way:

$$\langle \mathbf{E_1} \rangle = \frac{\mathbf{E_0}}{1-A_1\phi} \quad \text{with} \quad A_1 = \frac{\beta^0 + \beta^\infty X^2}{1+X^2} \qquad (16)$$

where, according to relation (8), X and $E_0$ are related by:

$$X=0, \quad E_0 < (1-\beta^0\phi)E_c$$

$$X^2 + 1 = \left(\frac{\langle E_1 \rangle}{E_c}\right)^2 = \frac{1}{(1-A_1\phi)^2}\left(\frac{E_0}{E_c}\right)^2, \quad E_0 > (1-\beta^0\phi)E_c \qquad (17)$$

Then, the current density in the suspension is determined upon using formulae (9) and (10) where $\langle \mathbf{E_2} \rangle$ and **P** are calculated assuming that each particle is submitted to $\langle \mathbf{E_1} \rangle$:

$$\gamma_s = \gamma_1 \left\{ 1 + 3\frac{A_1\phi}{1-A_1\phi} \right\} \qquad (18)$$

At first order in the particle volume fraction, this result simplifies to the expression (13) which has been obtained without accounting for the interaction between particles [6].

## 3. Experiments

### *3.1. Materials*

We have studied various suspensions of PMMA particles dispersed in several dielectric liquids. The PMMA particles provided by ICI Acrylics (Elvacite grade 2041) are sieved between 63 and 80 microns. Their dielectric constant is $2.6\varepsilon_0$ and their conductivity is about $10^{-14}$ S/m. Among the host liquids we used, we shall present some results concerning transformer oil (Dielec S, Hafa France) and a mixture of transformer oil and Ugilec (Elf Atochem) which is made in order to match the density of the particles ($\rho_2=1.2\ 10^3$kg.m$^{-3}$). The interested reader should find more results in [8]. The conductivity of the liquids is controlled by adding AOT salts (Sigma Aldrich, sodium dioctyl-sulfosuccinate). The characteristics of the suspending liquids are summarized in table 1.

| Liquid | Permittivity | [AOT] (mol/L) | Conductivity (S/m) | Viscosity (Pa.s.) | Critical field (V/mm) | Density (kg/m$^3$) |
|---|---|---|---|---|---|---|
| Dielec S | $2.4\varepsilon_0$ | 0.1 | $4.3\ 10^{-9}$ | $12.9\ 10^{-3}$ | 550 | $0.84\ 10^3$ |
| DielecS + Ugilec | $3.69\varepsilon_0$ | $10^{-3}$ | $3.3\ 10^{-8}$ | $13.6\ 10^{-3}$ | 1150 | $1.2\ 10^3$ |

Table 1

### *3.2. Device and experimental procedure*

The suspension is maintained by capillarity between two horizontal steel disks which have been hollowed to make ducts through which a thermoregulated fluid circulates. In this way, the temperature of the suspension is maintained to $25 \pm 0.5$°C. The lower electrode is raised on a jack. Its position is adjusted thanks to a micrometric screw and controlled with a precision of 10 microns thanks to a mechanical sensor. The diameter of the electrodes is 9 cm and the gap is usually set to 1 mm. A high voltage, V, comprised between 0 and $\pm 5000$V is applied across the electrodes. The voltage is supplied by using a Trek 1010B which amplifies (gain 1000) the signal generated by a PC. The electric current passing through the suspension is measured thanks to an electrometer (Keithley 617) and is recorded on a PC.

As we shall see in the following, the variations of the conductivity we expect to measure, due to Quincke rotation, are of the order of 10 or 15%. So the measurement of the current is to be done with a precision of about 1%. To reach this objective, we had to overcome several obstacles. The first one is the presence of a slight electric charge carried by the particles which is responsible for their agglutination on one electrode or the other. To get round this problem, the voltage polarity is periodically reversed with a characteristic time of 500 ms which is much larger than the Maxwell time (about few milliseconds) and much smaller than the particle electro-migration characteristic time (several tens of seconds).

A second obstacle arises from Joule heating which can be present even though the temperature of the electrodes is controlled thanks to a fluid circulation. To minimize its effect, we decided to apply the electric field by intermittence: the field-on periods are interspersed with field-off periods and, for each field intensity, the current is recorded at the end on the field-on period (see Figure 2).

A third difficulty comes from the complexity of the electrical properties of the host liquid itself. Indeed it is well known that Ohm's law is only of limited applicability for electrolytic resistances and that the conductance increases with the intensity of the field. This field conductivity enhancement is especially pronounced in low polar fluids and has been theoretically explained by Onsager [11]. Onsager theory is based on the effect of the electric field on the process of dissociation of ionic pairs: the dissociation constant, $k_D$, increases with the electric field intensity while the recombination one is unchanged:

$$\frac{\gamma_1(E_0)}{\gamma_1(0)} \propto \sqrt{\frac{k_d(E_0)}{k_d(0)}} = 2\frac{I_1(b^{\frac{1}{2}})}{b^{\frac{1}{2}}} \quad (19)$$

where $I_1$ is the first order modified Bessel function of the first kind and b is the ratio of the applied electric field on a characteristic value, $E_{dis}$, such that the electrostatic interaction between the anion and the cation in an ionic pair is of the same order of magnitude as the coulombic force exerted by the external field on the ions:

$$E_{dis} = \frac{\pi \varepsilon_1 (kT)^2}{e^3 E_0} \quad (20)$$

The confrontation between Onsager theory and experimental measurements of the field enhancement conductivity gives rise to contradictory results. Some authors report a rather good agreement [12] whereas others mention discrepancies [13]. With all the liquids we use (Dodecane, Dielec S and Ugilec with various salts), we note that the Onsager theory overestimates sensitively the conductance increase due to the electric field. An example of such a deviation from the Onsager theory is given in Figure 3. The Onsager's model applied with the theoretical value of $E_{dis}$, 260V/mm (curve a), leads to an important overestimation of the liquid conductivity whereas the experimental results are perfectly fitted with the Onsager's model using $E_{dis}$=325 V/mm (curve b). This observation needs further investigation to find an interpretation.

Nevertheless it can be seen that the host liquid conductivity can vary by more than 50% when the field intensity is raised up to 4000V/mm. Since our goal is to measure the effect of particle rotation on the apparent conductivity of the suspension, we have normalized the apparent conductivity of the suspension by the value of the host liquid conductivity, for each intensity of the electric field. It must be pointed out that the liquid is extracted from the suspension after particle centrifugation. Indeed, it is preferable to extract the liquid from the suspension than to use the liquid before adding the particles because, through chemical surface reactions, the presence of particles can modify the conductivity of the host liquid itself.

At last, we noted, like it is reported on Figure 4, curve (a), that the current passing through the liquid increased with time when a constant field intensity is applied. This problem does not exist anymore if the conductivity cell is placed under a nitrogen atmosphere (see Figure 4, curve (b)). This observation has been of central importance in our measurements since as soon as we began to make the experiments in a nitrogen atmosphere, the measurements became very reproducible. All the experimental results presented below have been obtained under such conditions and with the protocol described on Figure 2.

## 4. Experimental results and discussion

The first experiments have been carried out with PMMA particles dispersed into Dielec S for two different particle volume fractions: 5% and 10%. The apparent conductivity of these suspensions is reported on Figures 5 and 6. The solid lines represent the results of the model presented before and the dotted ones correspond to the first order approximation in $\phi$. The symbols represent the experimental results: the triangles correspond to an experiment where the electric field has been raised from zero to its maximum value while the squares have been obtained by decreasing the field intensity. For both particle volume fractions, the qualitative behaviour of the conductivity above the critical field is satisfactory since a quite abrupt increase of the apparent conductivity is observed just above the critical field and a high conductivity plateau is present at high electric field intensity.

For the lowest volume fraction, the numerical agreement between the model predictions and the experimental results is good while for the highest particle concentration, the model underestimates the conductivity. The value of the particle volume fraction itself and so the electrostatic or hydrodynamic interactions between particles cannot be invoked to explain this discrepancy since, as we will see in the following, we have obtained other results for such concentrated suspensions which, above the critical field, match very accurately the theoretical predictions. For both particle concentrations, the experimental and theoretical values of the critical electric field are close together. Below the critical field, the experimental behaviour is far from the expected one, especially for the experiments made increasing the field intensity. Undoubtedly, this mismatch originates in the sedimentation of the particles which form a poor conducting layer on the lower electrode giving rise to a low apparent conductivity. When the field is applied, the particles are polarised and attract each other in the field direction so that they pile up on each other to form chains. The layer decompaction which accompanies the particle chaining is responsible for a continuous increase of the apparent conductivity. This argument is also convenient for explaining why we measure a different value of the apparent conductivity whether the field intensity is increased or decreased. Indeed, in the first case, all the particles are settled at the bottom while, in the second case, when the field is decreased to the critical filed, the particles are homogeneously distributed thanks to their rotation which acts like a resuspension force. When the field becomes lower than the critical field, particles form chains which connect one electrode to the other. When the field is decreased further, the dipolar electric forces which stabilise the chain structure decrease and, below a given field intensity, $E_S$, they are dominated by the gravity forces and the chains break. An order of magnitude for $E_S$ can be obtained by equating the dipolar and gravitational forces:

$$F_{dip} \approx \frac{3}{8}\pi\varepsilon_1 a^2 E^2$$

$$F_{grav} \approx \frac{4}{3}\pi a^3 (\rho_2 - \rho_1) g$$

(21)

This comparison gives $E_S \approx 150$ V/mm which is not so far from the experimental observation (see Figure 5 or 6).

Above this field intensity, the suspension is assumed to structure itself in dense columnar aggregates whose particle volume fraction is $\phi_a$. $\phi_a$ can be evaluated upon assuming that the particle organisation within an aggregate is the same as for a classical electro-rheological suspension [14]. In such a case, in the aggregates, the particles are distributed on a body-centered-tetragonal array whose compacity is $\phi_a = 0.68$ and the suspension consists of two phases, the aggregates which occupy a volume fraction of $\phi/\phi_a$ and the suspending liquid. Therefore, the effective conductivity of the suspension is obtained by computing the weighted average of the conductivity over the aggregates and the clear liquid. Upon using expression (18) to calculate the conductivity inside an aggregate, the average conductivity of the structured suspension is found to be:

$$\gamma_s^{struc} = \gamma_1 \left\{ 1 + \frac{3\beta^0 \phi}{1 - \beta^0 \phi_a} \right\}$$

(22)

This value of the conductivity is represented on Figures 5 and 6 by the dashed lines and is in rather good accordance with the experimental measurements.

Furthermore the organisation of the suspension in dense aggregates is likely to be responsible for an increase of the critical field intensity since the local field in the liquid decreases with the particle volume fraction. Consequently, according to the equation (16), the particles that belong to an aggregate are expected to rotate only if the external field intensity is higher than:

$$E_c^{struc} = \left(1 - \beta^0 \phi_a\right) E_c \qquad (23)$$

Moreover, it means that the measured critical field intensity is expected to be higher when the electric field is increased and the particles form aggregates than when it is decreased and the particles are randomly distributed. This difference does not appear clearly on Figures 5 and 6, probably because, due to the density mismatch between the particles and the liquid, bellow the critical field, the aggregates have not completely formed.

On the opposite, the hysteresis is noticeable on Figure 7 which concerns the conductivity of a neutrally buoyant suspension of PMMA particles dispersed in a mixture of Dielec S and Ugilec. On this figure, the triangles and the squares represent the conductivity that has been measured upon increasing and decreasing the field respectively. Below the critical field, the structuring of the suspension manifests itself by the increase of the conductivity from a low value which is close to the expected one for a homogeneous suspension (expected normalised conductivity 0.86, measured conductivity 0.87) to a higher conductivity value expected for a structured suspension (expected normalised conductivity 0.89, measured conductivity 0.92).

On Figure 7, the solid line represents the theoretical conductivity of the suspension when the particles are homogeneously distributed and the dashed line accounts for the modification of both the conductivity below the critical field and the critical field value itself due to the organisation of the suspension in dense aggregates. Although the hysteresis area is roughly the same experimentally and theoretically, it is worth noting that the above model underestimates significantly the minimum electric field intensity for the particles to rotate. This discrepancy could of course originate in the role played by the hydrodynamic forces between particles which are not accounted for in our model.

## 5. Conclusion

We have studied the influence of Quincke rotation on the conductivity of a suspension. Both theoretically and experimentally, we have demonstrated that the particle rotation was responsible for an increase of the order of 15% in the suspension conductivity. This conductivity enhancement is explained by the increase of electric charge transport by convection when the particles rotate.

Otherwise, the necessity to measure very accurately the conductivities of the suspension and of the host liquid has required using a strict operating procedure. In particular, the importance of making the conductivity measurements under an inert atmosphere has been addressed.

## Acknowledgements

We acknowledge A. Audoly for his technical support

# Figure caption

Figure 1: If the charge relaxation time is higher in the particle than in the liquid, the charge which accumulates at the particle/liquid interface is equivalent to that of a dipole moment whose direction is opposite to the electric field (a). If a sufficiently high field intensity is applied, this situation is unstable and the particle rotates.

Figure 2: Presentation of the experimental procedure used to measure the current passing through the suspension.

Figure 3: An illustration of electricfield-enhanced conductivity of Dielec S with AOT ($10^{-1}$ mol/L). The stars represent the experimental results. The curve (a) stands for Onsager theory ($E_{dis}$=260V/mm) and the curve (b) is a fit of the experimental data with Onsager expression where the characteristic electric field $E_{dis}$ has been modified ($E_{dis}$=325V/mm).

Figure 4: Comparison of the current density passing through the measuring cell (surface 64cm$^2$) filled with Dielec S and AOT ($10^{-1}$mol/L) when placed in air (a) or in nitrogen atmosphere (b). The electric field intensity is 4000V/mm. The difference in the current densities recorded at t=0 originates in a slight difference in the AOT concentration but does not deal with the experimental technique.

Figure 5: Apparent conductivity of a suspension of PMMA particles ($\phi$=5%) dispersed in Dielec S containing AOT ($10^{-1}$ mol/L). Symbols: experimental results obtained either by increasing the electric field intensity (triangles) or by decreasing it (squares). Solid curve: model prediction, dotted line: first approximation in particle volume fraction, dashed line: prediction of the conductivity of the suspension when it is structured in columnar aggregates.

Figure 6: Same as Figure 5 except that the particle volume fraction is $\phi$=10%

Figure 7: Apparent conductivity of a suspension of PMMA particles ($\phi$=10%) dispersed in a mixture of Dielec S (32% in mass) and Ugilec (68% in mass) whose density is that of the particles. The liquid contains AOT ($10^{-3}$ mol/L). Solid curve: model prediction. Dotted line: first approximation in particle volume fraction. Dashed line: prediction of the conductivity of the suspension when structured in columnar aggregates.

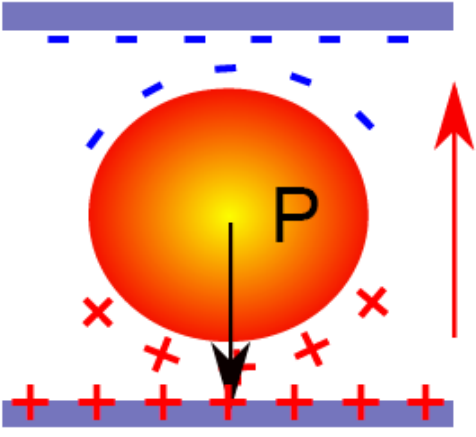 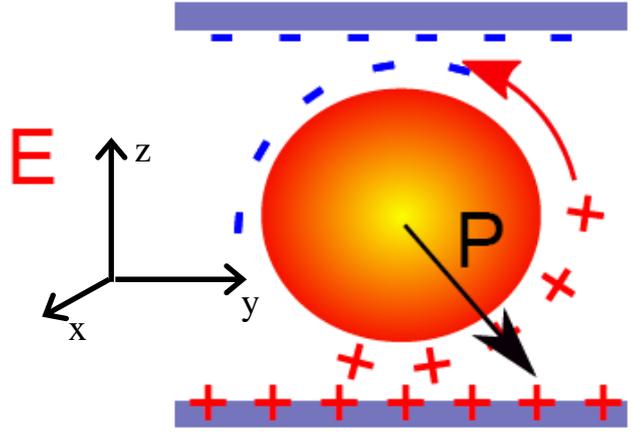

Figure 1.a                                                                 Figure 1.b

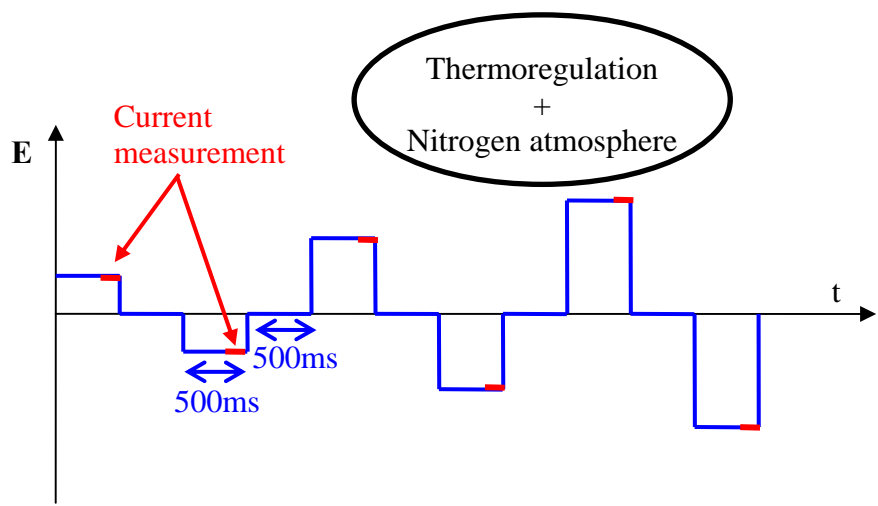

Figure 2

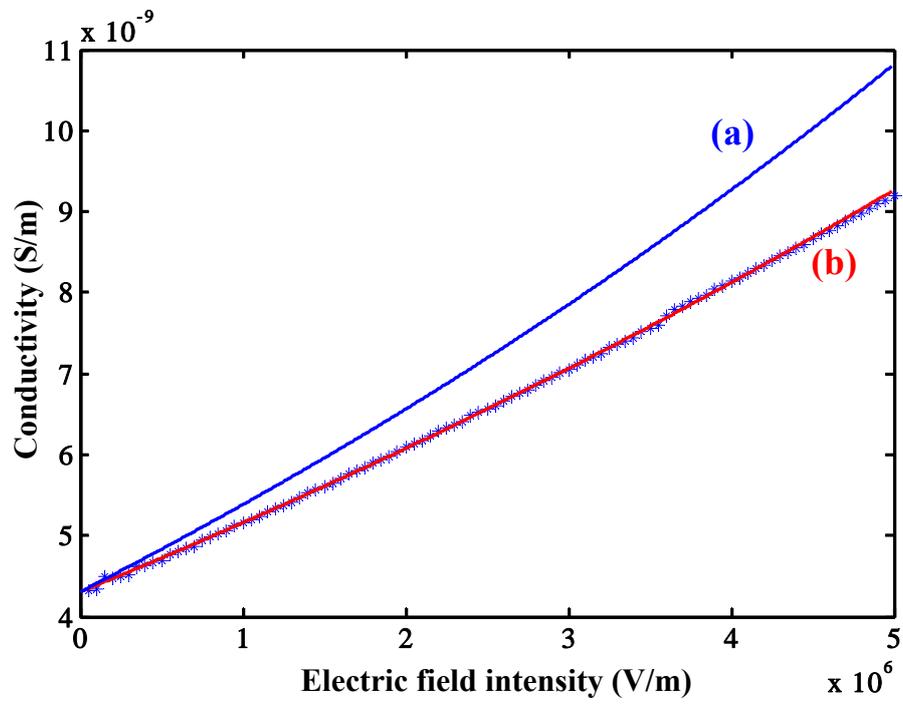

Figure 3

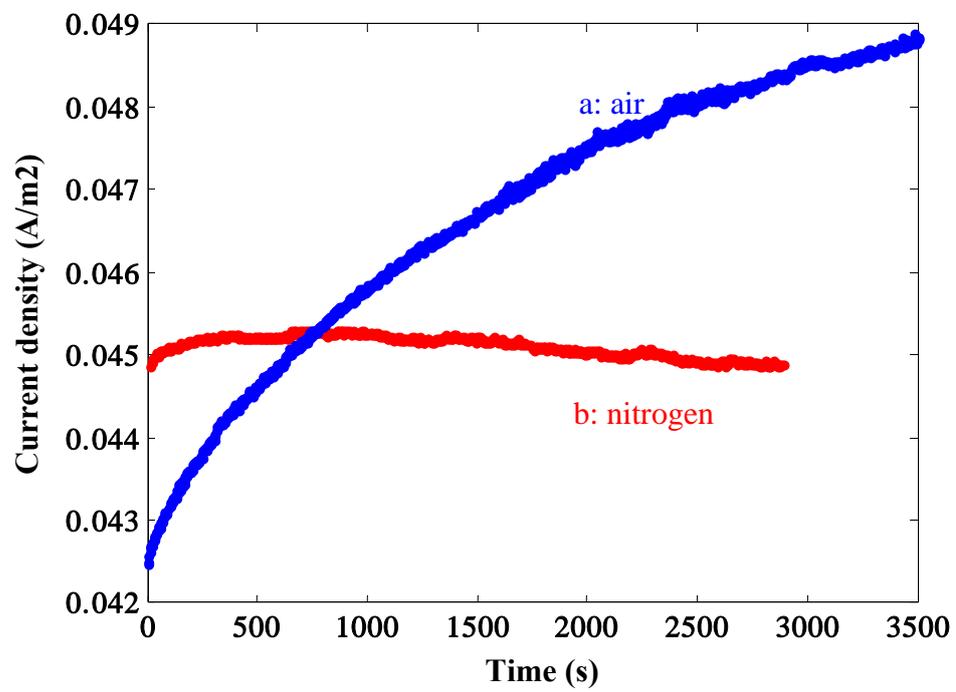

Figure 4

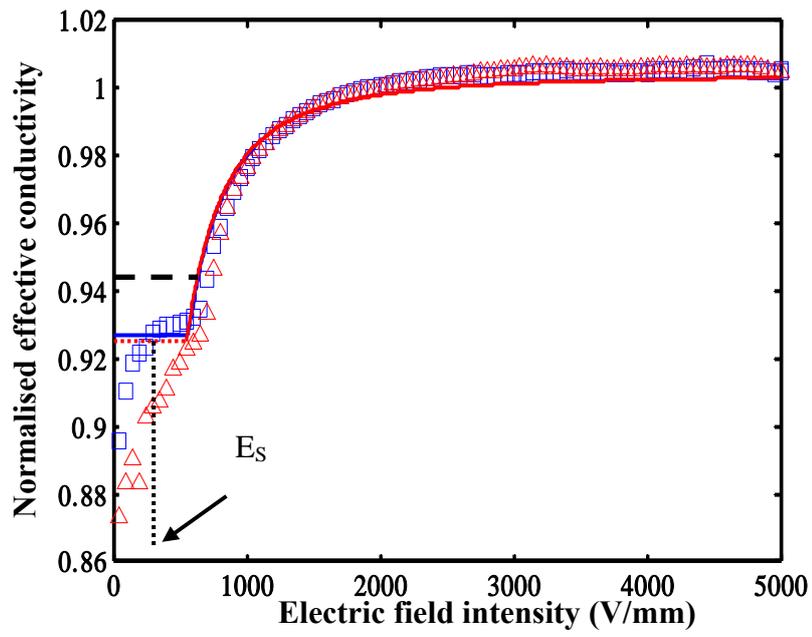

Figure 5

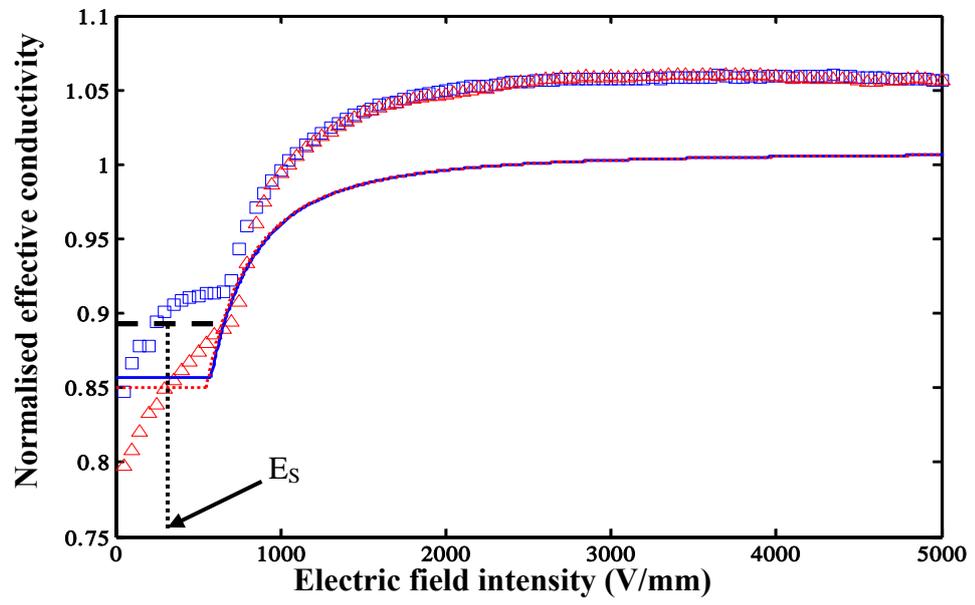

Figure 6

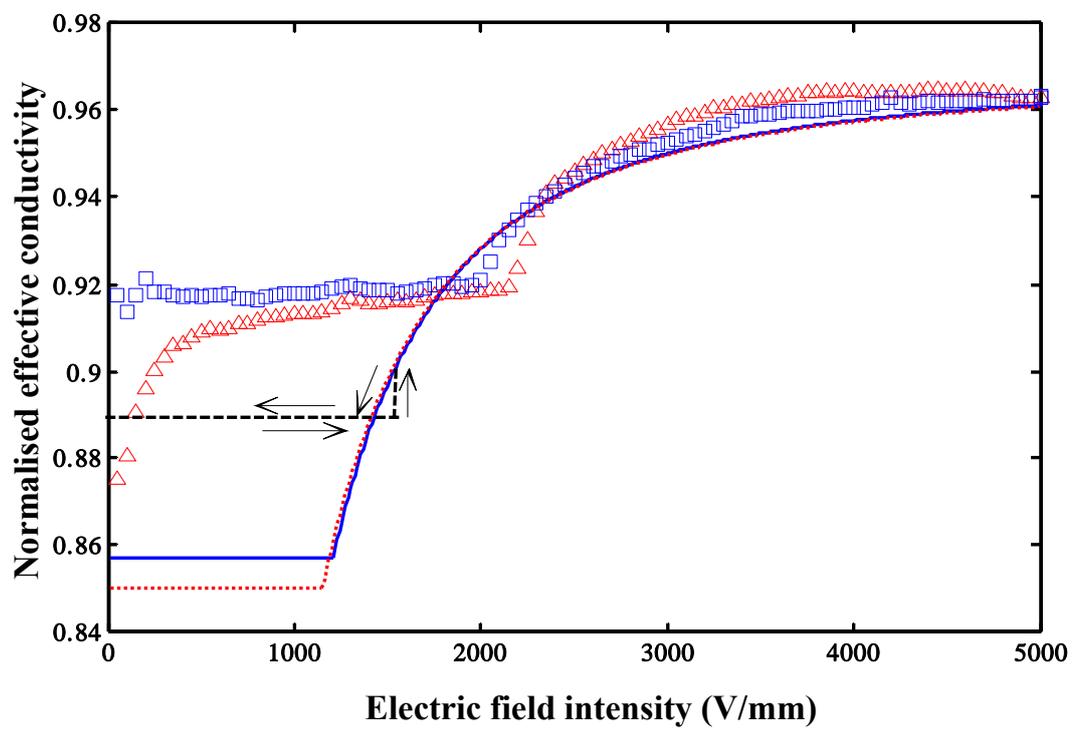

Figure 7